\documentclass[12pt]{article}

\usepackage{a4}
\usepackage{epsfig}

\newlength{\abstwidth}
\def\be{\begin{equation}}
\def\ee{\end{equation}}

\begin{document}

\def\lsim{\mathrel{\rlap{\lower4pt\hbox{\hskip1pt$\sim$}}
    \raise1pt\hbox{$<$}}}         
\def\gsim{\mathrel{\rlap{\lower4pt\hbox{\hskip1pt$\sim$}}
    \raise1pt\hbox{$>$}}}         

\pagestyle{empty}

\begin{flushright}
BI-TH 2002/04\\
\end{flushright}

\vspace{\fill}

\begin{center}
{\Large\bf The longitudinal structure function of the proton for small x
${}^{*}$}
\\[1.8ex]
{\bf Dieter Schildknecht} \\[1.2mm]
Fakult\"{a}t f\"{u}r Physik, Universit\"{a}t Bielefeld \\[1.2mm]
D-33501 Bielefeld, Germany \\[1.5ex]
and \\[1.5ex]
{\bf Mikhail Tentyukov${}^{**}$} \\[1.2mm]
Fakult\"{a}t f\"{u}r Physik, Universit\"{a}t Bielefeld \\[1.2mm]
D-33501 Bielefeld, Germany
\end{center}

\vspace{\fill}

\begin{center}
{\bf Abstract}\\[2ex]
\begin{minipage}{\abstwidth}
A comparison of the H1 data on the longitudinal structure function, $F_L$,
at small $x$ with the predictions from
the generalized vector dominance / color dipole picture (GVD/CDP) is
presented. Using the set of parameters previously determined in the fits to
the total cross section, $\sigma_{\gamma^* p}$, we find good agreement with the
data for $F_L$. Scaling in $\eta = (Q^2 + m^2_0 ) / \Lambda^2 (W^2)$ is
discussed in detail for the longitudinal and transverse photoabsorption cross
sections.
\end{minipage}
\end{center}

\vspace{\fill}
\noindent

\rule{60mm}{0.4mm}

\vspace{0.1mm}
\noindent
${}^*$ Supported by the BMBF, Contract 05 HT9PBA2\\
${}^{**}$ On leave from BLTP JINR, Dubna, Russia
\clearpage
\pagestyle{plain}
\setcounter{page}{1}

\baselineskip 20pt

A confrontation of predictions for polarisation phenomena with experimental
data has frequently in the past provided crucial tests of a theoretical
ansatz. This was also true for an analysis of the nucleon structure functions
when
discriminating between longitudinal and transverse virtual photons.

The H1 collaboration at HERA has recently presented data \cite{1,2}
for the longitudinal structure function of the proton, $F_L (x, Q^2)$,
in the diffraction region of small valus of $x \cong Q^2 / W^2 \ll 1$.
In the present note, we compare the experimental data with the predictions
from the generalized vector dominance/color dipole picture (GVD/CDP)
\cite{3,4,5}\footnote{For somewhat related approaches compare
refs.\cite{6,7}}.
We will present our analysis of the data in terms of the longitudinal
structure function, $F_L (x , Q^2)$, and in terms of the longitudinal part of
the photoabsorption cross section, $\sigma_L (W^2 , Q^2)$. In the case of
the longitudinal cross section, $\sigma_L (W^2 , Q^2)$, we will employ the
scaling variable
\be
\eta (W^2 , Q^2) = \frac{Q^2 + m^2_0}{\Lambda^2 (W^2)} ,
\label{(1)}
\ee
recently introduced \cite{3,4} in our analysis of the total cross section,
where in good approximation
\begin{eqnarray}
\sigma_{\gamma^* p} (W^2 , Q^2)& =& \sigma_T (W^2 , Q^2) + \sigma_L (W^2 , Q^2)
\nonumber \\
& = & \sigma_{\gamma^* p} (\eta (W^2 , Q^2)) .
\label{(2)}
\end{eqnarray}
Here $\Lambda^2 (W^2)$ is a slowly increasing function of $W^2$ proportional
to the effective gluon momentum transfer squared, and $m_0$ denotes
a threshold mass.
We will provide a detailed discussion of the behavior of $\sigma_T$ and
$\sigma_L$ with respect to their dependence on $\eta$, in particular in the
$Q^2 \rightarrow 0$ photoproduction limit, where $\eta \rightarrow
\eta_{{\rm Min}} = m^2_0 / \Lambda^2 (W^2)$.

The GVD/CDP was described and compared with the experimental data for the
total cross section, $\sigma_{\gamma^* p}$, in refs.\cite{3,4,5}.
The GVD/CDP, equivalently, may be formulated in transverse position space
or in momentum space.

The evaluation of the two-gluon exchange diagrams \cite{7a} in the low-$x$ limit, upon
transition to transverse position space, leads to the representation \cite{8}
\be
\sigma_{\gamma^* p} (W^2 , Q^2) = \int^1_0 \, dz \, \int \, d^2 r_\perp
| \psi |^2 (\vec r^{~2}_\perp , z (1-z), Q^2) \sigma_{(q \bar q) p} (
\vec r_\perp^{~2}, z (1-z), W^2),
\label{(3)}
\ee
where $\psi$ denotes the so-called photon wave function, explicitly given
in \cite{8}. The color-dipole cross section, $\sigma_{(q \bar q)p}$,
fulfills a Fourier representation of the form
\be
\sigma_{(q \bar q)p} (\vec r^{~2}_\perp , z (1 - z) , W^2) = \int d^2 l_\perp
\tilde\sigma_{(q \bar q)p} (\vec l^{~2}_\perp , z (1 - z), W^2)\left( 1 - e^
{-i \vec l_\perp \cdot \vec r_\perp} \right) .
\label{(4)}
\ee
In (\ref{(3)}) and (\ref{(4)}), the variables $\vec r_\perp$ and $z$ denote
the two-dimensional vector of the transverse interquark separation and the
fraction of the photon momentum taken over by one of the incoming quarks.
The representation (\ref{(3)}), with (\ref{(4)}), contains the underlying
generic structure of two-gluon exchange, and, accordingly, it encorporates
``color transparency'' as well as hadronic unitarity provided appropriate
convergence properties are readily fulfilled. The empirical scaling
behavior (\ref{(2)}) (compare \cite{3,4}) is embodied in (\ref{(3)}) by
adopting the simple ansatz of
\be
\tilde\sigma_{(q \bar q)p} (\vec l^{~2}_\perp , z (1 - z), W^2) =
\sigma^{(\infty)} \frac{1}{\pi} \delta (\vec l^{~2}_\perp - \Lambda^2 (W^2) z
(1-z))
\label{(5)}
\ee
for the gluon-gluon-proton-proton vertex function,
$\tilde\sigma_{(q \bar q)p}$, in (\ref{(4)}). Substitution of (\ref{(5)}) into
(\ref{(4)}) yields
\be
\sigma_{(q \bar q)p} (\vec r^{~2}_\perp , z (1 - z) , W^2) =
\sigma^{(\infty)} (1 - J_0 (r_\perp \sqrt{z(1-z)} \Lambda (W^2))) ,
\label{(6)}
\ee
where $J_0$ denotes a Bessel function. The asymptotic value of the dipole cross
section for $r_\perp \rightarrow \infty$ as well as $W^2
\rightarrow \infty$ has been denoted by $\sigma^{(\infty)}$. Actually,
$\sigma^{(\infty)}$ is constant in good approximation, and it is of typical
hadronic magnitude. Equations (\ref{(3)}) and (\ref{(6)}) may be
considered as the basic formulae of the GVD/CDP. They yield $\sigma_{\gamma^*
p} = \sigma_{\gamma^* p} (\eta)$, and accordingly, $\sigma_{\gamma^* p}$
depends on the (threshold) mass
$m^2_0$ and on the adjustable parameters describing the increase of
of the average or effective gluon transverse momentum, $\Lambda^2 (W^2)$,
with energy.

Actually, the evaluation of the GVD/CDP was carried out \cite{9,3}
in momentum space. Inserting (\ref{(4)}) into (\ref{(3)}) together with the
Fourier representation of the photon wave function for longitudinal
and transverse (virtual) photons takes us back to momentum space and leads to
\be
\sigma_{\gamma^*_{T,L} p} (W^2 , Q^2) = \frac{\alpha R_{e^+ e^-}}{3\pi}
\sigma^{(\infty)} I_{T,L} \left( \eta , \frac{m^2_0}{\Lambda^2 (W^2)}
\right)
\label{(7)}
\ee
with
\be
R_{e^+ e^-} = 3 \sum Q^2_i ,
\label{(8)}
\ee
where $R_{e^+ e^-} = 2$ is to be inserted, since upon specifying
(\ref{(7)}) to
photoproduction, $Q^2 = 0$, only three flavors $(u, d, s)$ with
charges $Q_i$ contribute appreciably, while $\sigma^{(\infty)}\cong 80
GeV^{-2}\cong 31 mb$ {~~}\footnote{As long as no detailed treatment of the
influence of the charm-quark mass is included in (\ref{(7)}), the dependence
on the product $R_{e^+ e^-} \cdot \sigma^{(\infty)}$ allows one to equally
well insert $R_{e^+ e^-}=10/3$ and $\sigma^{(\infty)}=48 GeV^{-2} = 18.7 mb$}.
The (dimensionless) functions $I_{T,L} (\eta , m^2_0 / \Lambda^2 (W^2))$
in (\ref{(7)}) denote integrals of the form of mass dispersion relations
reminiscent of off-diagonal generalized vector dominance \cite{10,11}.
They were represented \cite{3} as a sum of two terms,
\be
I_{T,L} = I_{T,L}^{(1)} + I_{T,L}^{(2)} .
\label{(9)}
\ee
The main transverse and longitudinal, contributions $I_{T,L}^{(1)}$, are given
by
\begin{eqnarray}
& & I_T^{(1)} \left( \eta , \frac{m^2_0}{\Lambda^2 (W^2)} \right)  \nonumber \\
& & = \frac{1}{\pi} \int^\infty_{m^2_0} \, dM^2 \int^{(M+\Lambda (W^2))^2}_
{(M-\Lambda (W^2))^2} dM^{\prime 2} \omega (M^2 , M^{\prime 2} , \Lambda^2
(W^2)) \label{(10)} \\
& & \times \left[ \frac{M^2}{(Q^2 + M^2)^2} - \frac{M^{\prime 2} + M^2 -
\Lambda^2 (W^2)}{2 (Q^2 + M^2)(Q^2 + M^{\prime 2})} \right] , \nonumber
\end{eqnarray}
and
\begin{eqnarray}
& & I_L^{(1)} \left( \eta , \frac{m^2_0}{\Lambda^2 (W^2)} \right)  \nonumber \\
& & = \frac{1}{\pi} \int^\infty_{m^2_0} \, dM^2 \int^{(M+\Lambda (W^2))^2}_
{(M-\Lambda (W^2))^2} dM^{\prime 2} \omega (M^2 , M^{\prime 2} , \Lambda^2
(W^2))  \label{(11)} \\
& & \times \left[ \frac{Q^2}{(Q^2 + M^2)^2} - \frac{Q^2}
{2 (Q^2 + M^2)(Q^2 + M^{\prime 2})} \right] . \nonumber
\end{eqnarray}
The terms $I_{T,L}^{(2)}$ in (\ref{(9)}) assure the correct threshold behavior
of $I_{T,L}$ in the off-diagonal contribution\footnote{Numerically, it turns
out that $I^{(2)}_T$ is practically negligible in the HERA energy range, while
$I^{(2)}_L$ contributes about 20\% at the lowest HERA energy and becomes
negligible, when the energy reaches the highest HERA energy. Note that
explicit analytic formulae are available \cite{3} for $I_{T,L}^{(1)}$, and
accordingly only the correction term $I_{L}^{(2)}$ must be evaluated by
numerical integration.}.
They are given by
\begin{eqnarray}
& & I_T^{(2)} \left( \eta , \frac{m^2_0}{\Lambda^2 (W^2)} \right)  \nonumber \\
& & = \frac{1}{\pi} \int^\infty_{m^2_0} dM^2 \Theta (m^2_0 - (M - \Lambda
(W^2))^2) \int^{m^2_0}_{(M-\Lambda(W^2))^2} dM^{\prime 2} \\
& & \times \omega (M^2 , M^{\prime 2} , \Lambda^2 (W^2))
\frac{M^{\prime 2} + M^2 - \Lambda^2 (W^2)}{2 (Q^2 + M^2)(Q^2 + M^{\prime 2})}
,\nonumber
\label{(12)}
\end{eqnarray}
and
\begin{eqnarray}
& & I_L^{(2)} =  \left( \eta , \frac{m^2_0}{\Lambda^2 (W^2)} \right)\nonumber\\
& & = \frac{1}{\pi} \int^\infty_{m^2_0} dM^2 \Theta (m^2_0 - (M - \Lambda
(W^2))^2) \int^{m^2_0}_{(M-\Lambda(W^2))^2} dM^{\prime 2} \\
& & \times \omega (M^2 , M^{\prime 2} , \Lambda^2 (W^2)) \frac{Q^2}{(Q^2 + M^2
) (Q^2 + M^{\prime 2})}. \nonumber
\label{(13)}
\end{eqnarray}
We note the relative minus sign between the diagonal $(M^2 = M^{\prime 2})$
and the
off-diagonal propagator term in (\ref{(10)}) and (\ref{(11)})
that is characteristic
for off-diagonal generalized vector dominance \cite{10}.
The two contributions with their relative minus sign are an outgrowth of
the two-gluon exchange structure.
Concerning $\omega (M^2, M^{\prime 2}, \Lambda^2 (W^2))$, we only note the
integration formulae,
\be
\frac{1}{\pi} \int^{(M+\Lambda(W^2))^2}_{(M-\Lambda(W^2))^2} dM^{\prime 2}
\omega (M^2, M^{\prime 2}, \Lambda^2 (W^2)) = 1,
\label{(2.18)}
\ee
and
\be
\frac{1}{\pi} \int^{(M+\Lambda(W^2))^2}_{(M-\Lambda(W^2))^2} dM^{\prime 2}
\omega (M^2, M^{\prime 2}, \Lambda^2 (W^2)) M^{\prime 2} = M^2 + \Lambda^2
(W^2),
\label{(2.19)}
\ee
and refer to ref.~\cite{3} for the explicit expressions.

In \cite{3}, we gave explicit analytical expressions for $I_{T,L}^{(1)}$,
and an approximation formula for the sum $I_T^{(2)} + I_L^{(2)}$.
These formulae allowed us to perform a fit to the data for the total cross
section, $\sigma_{\gamma^* p} (W^2 , Q^2)$. The fit gave the parameters
\be
m^2_0 = 0.16 \pm 0.01 \, {\rm GeV}^2 ,
\label{(14)}
\ee
as well as the parameters describing the increase of $\Lambda^2 (W^2)$
with energy, alternatively by a power law or a logarithm,
\be
\Lambda^2 (W^2) = \left\{ \matrix{ & C_1 (W^2 + W^2_0)^{C_2} , \cr
& C^\prime_1 \ln \left( \frac{W^2}{W^2_0} + C^\prime_2 \right) , }
\right.
\label{(15)}
\ee
where
\begin{eqnarray}
C_1 & = & 0.34 \pm 0.05 ({\rm GeV}^2)^{1-C_2} , \nonumber  \\
C_2 & = & 0.27 \pm 0.01 , \label{(16)} \\
W^2_0 & = & 882 \pm 246 {\rm GeV}^2 , \nonumber
\end{eqnarray}
and

\begin{figure}[h]
\begin{center}
{\centerline{\epsfig{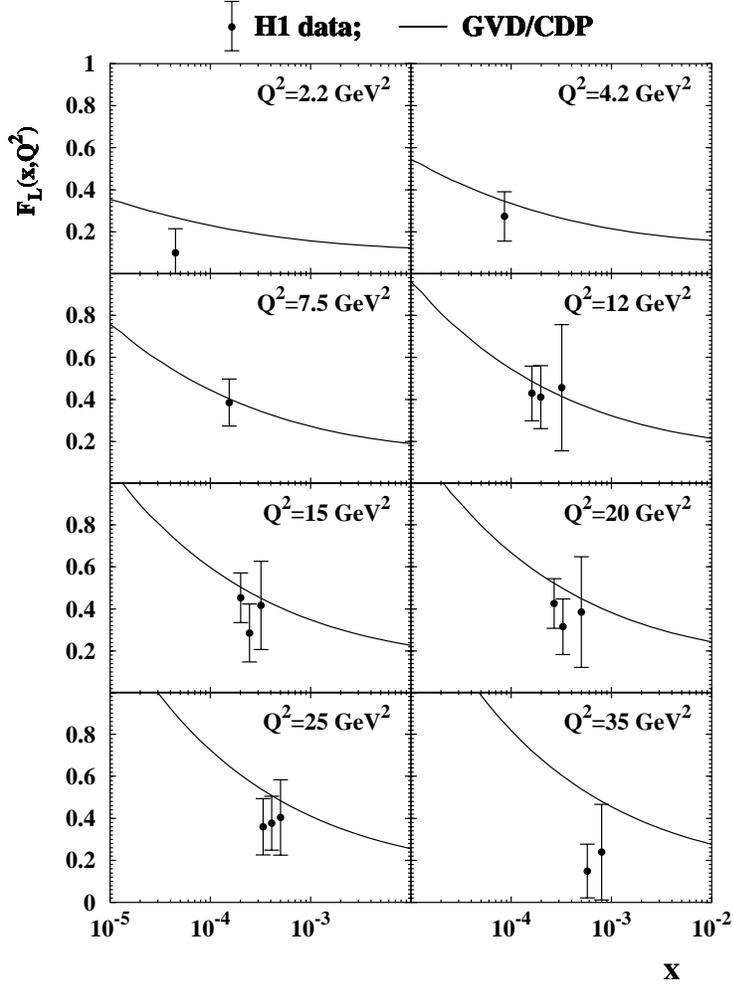}}}
\label{fig1}
\caption{
The H1 data for $F_L$ are compared with the predictions from the GVD/CDP.
}
\end{center}
\end{figure}

\noindent 
\begin{eqnarray}
C_1^\prime & = & 1.64 \pm 0.14 ({\rm GeV}^2) , \nonumber  \\
C_2^\prime & = & 4.1 \pm 0.4 , \label{(17)} \\
W^{\prime 2}_0 & = & 1015 \pm 334 {\rm GeV}^2 . \nonumber
\end{eqnarray}
Note that $\Lambda^2 (W^2)$ varies between $\Lambda^2 \simeq 2 {\rm GeV}^2$
and $\Lambda^2 \simeq 7 {\rm GeV}^2$ in the HERA energy range of $W^2 \simeq
1000 {\rm GeV}^2$ to $90000 {\rm GeV}^2$. Averaging over the
configuration variable $z$ yields $\langle \vec l^{~2}_\perp \rangle_{W^2} =
(1/6) \Lambda^2$ \cite{3}, i.e. a reasonable value of
$\langle \vec l^{~2}_\perp
\rangle \simeq 0.3 {\rm GeV}^2$ to  $\langle \vec l^{~2}_\perp
\rangle \simeq 1 {\rm GeV}^2$ for the average or effective gluon transverse
momentum squared absorbed by one of the quarks.

In fig. 1, we compare our results for
\be
F_L (x , Q^2) = \frac{Q^2}{4\pi^2 \alpha} \sigma_L (x , Q^2), \,\,\,
(x \ll 1),
\label{(18)}
\ee
with the H1 data using the parameters (\ref{(14)}) as well as (\ref{(16)})
or (\ref{(17)}) previously determined in the fits to $\sigma_{\gamma^* p}$.
There is reasonable agreement with the experimental data.

\begin{figure}[ht]
\begin{center}
{\centerline{\epsfig{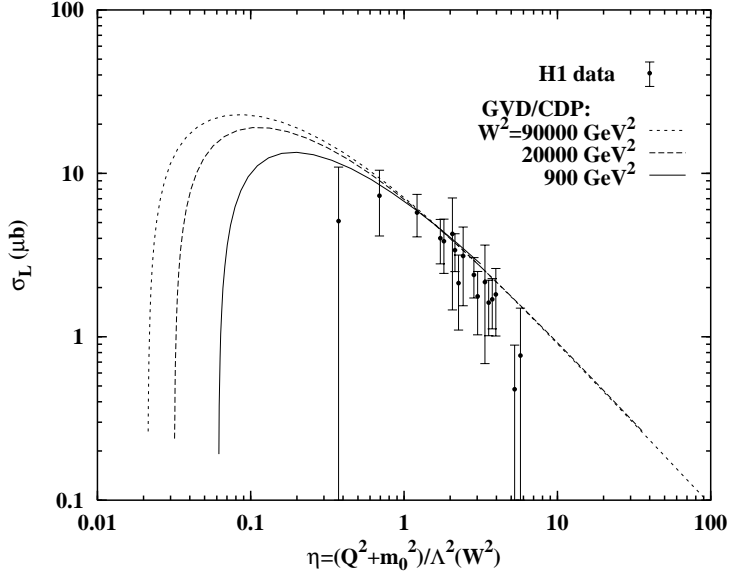}}}
\label{fig2}
\caption{
The H1 data, expressed in terms of $\sigma_L$, are compared with the
predictions from the GVD/CDP.
}
\end{center}
\end{figure}

From the point of view of the GVD/CDP it is more appropriate to plot the data
in terms of the longitudinal part of the total cross section. In fig. 2,
we show the data for $\sigma_L$, calculated from the data for $F_L$
according to (\ref{(18)}), plotted against the scaling variable $\eta$.
As expected from fig.~1, there is reasonable agreement, with a tendency for
the data to lie somewhat lower than the theoretical predictions, as also
seen in fig.~1.

The lower limit of $\eta$ determined by the photoproduction limit of
$Q^2 = 0$,
\be
\eta \ge \eta_{{\rm Min}} (Q^2 = 0, W^2) = \frac{m^2_0}{\Lambda^2 (W^2)} ,
\label{(19)}
\ee
decreases with increasing energy. The vanishing of $\sigma_L (\eta ,
\eta_{{\rm Min}})$ with $\eta - \eta_{{\rm Min}} = Q^2 / \Lambda^2 (W^2)
\rightarrow 0$ as a function of $\eta$, accordingly, occurs at values of
$\eta_{{\rm Min}}$ that decrease with increasing energy. Compare fig. 2.
In the low $Q^2$ regime of $\eta \cong \eta_{{\rm Min}}$, the longitudinal
cross section, $\sigma_L$, strongly depends on $\eta_{{\rm Min}}$ at
fixed $\eta$.
Scaling in $\eta$ is strongly violated. This is in contrast, as
previously discussed \cite{3}, to the total cross section, where
$\sigma_{\gamma^* p} = \sigma_{\gamma^* p} (\eta)$.

\begin{figure}[h]
\begin{center}
{\centerline{\epsfig{file=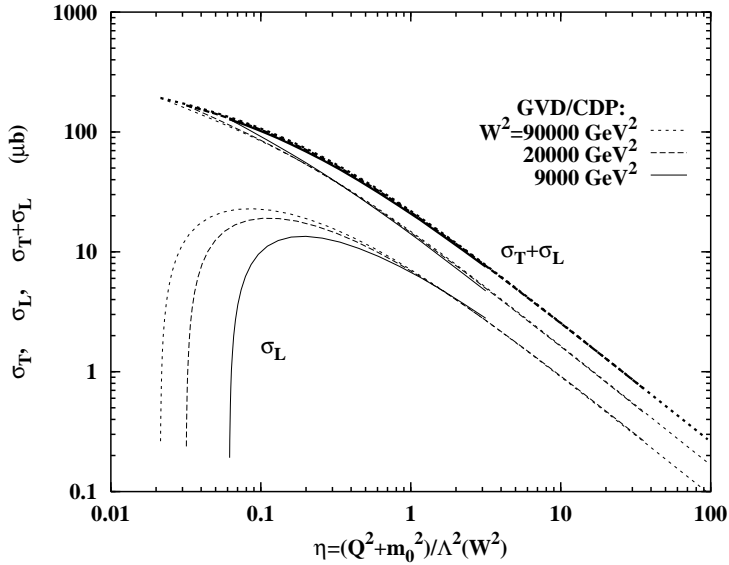,width=10.0cm}}}
\label{fig3}
\caption{
The theoretical predictions of the GVD/CDP for
$\sigma_L , \sigma_T$ and $\sigma_{\gamma^* p} = \sigma_L + \sigma_T$.
Note that the validity of the theoretical
predictions is restricted by $x\lsim 0.01$.
This restriction has been imposed on the curves in the figure.
}
\end{center}
\end{figure}

The behavior of $\sigma_T , \sigma_L$ as well as
$\sigma_{\gamma^* p} = \sigma_T + \sigma_L$ as a function of $\eta$
is shown in fig. 3. In order
to illuminate the behavior of the different contributions to the total
cross section, we note that $I^{(1)}_T$ and $I^{(1)}_L$ from (\ref{(10)})
and (\ref{(11)}) depend on $\eta$ and $\eta_{{\rm Min}}$,
\be
I_{T,L} = I_{T,L} (\eta , \eta_{{\rm Min}} ) .
\label{(20)}
\ee
Explicit formulae
were given in ref.~\cite{3}\footnote{Compare Appendix B in Eur. Phys. J
C20 (2001) 77 \cite{3}}.
Numerically, $\eta_{{\rm Min}} < 0.1$. Expressing $I_{T,L}$ in (\ref{(20)})
in terms of $\eta$ and
\be
\epsilon = \eta - \eta_{{\rm Min}} ,
\label{(21)}
\ee
by substitution of $\eta_{{\rm Min}} = \eta - \epsilon$, and noting that
$\epsilon \cong \eta$ as soon as $\eta > 1$, we immediately see that $I_{T,L}
= I_{T,L} (\eta)$ for $\eta > 1$; we have scaling in $\eta$ for $\sigma_T$
as well as $\sigma_L$ for $\eta > 1$. We turn to $\eta \cong \eta_{{\rm Min}}
< 1$. From the analytic expressions given in Appendix B of ref.~\cite{3},
one immediately notes that $I_T$ contains an additive contribution, opposite in
sign, but equal in magnitude to $I_L$. The violent increase of $I_L$ at
fixed $\eta$ with decreasing $\eta_{{\rm Min}}$ for $\eta \cong
\eta_{{\rm Min}} < 1$
is indeed seen to be entirely
cancelled, once the sum of $I_T$ and $I_L$ is taken. An expansion of the sum
of $I_T$ and $I_L$ at fixed $\eta$ as a function of $\eta_{{\rm Min}}$ shows
that any additional dependence on $\eta_{{\rm Min}}$ is negligible \cite{3}.

In summary, we have shown that the GVD/CDP with the parameters previously
determined in a fit to $\sigma_{\gamma^* p}$ describes the
longitudinal structure function, $F_L$, or, equivalently, the longitudinal
photoabsorption cross section, $\sigma_L$, at low
$x$. We have given a detailed discussion on how scaling in $\eta$ for
$\sigma_{\gamma^* p}$, i.e. $\sigma_{\gamma^* p} = \sigma_{\gamma^* p}
(\eta)$, arises despite the fact that scaling is strongly violated for the
longitudinal cross section in the region of $\eta \cong \eta_{{\rm Min}}$.

\bigskip\noindent
{\bf Acknowledgement}

It is a pleasure to thank Nikolai Nikolaev for useful discussions.

\end{document}